\documentclass[final,5p,times,twocolumn,nopreprintline]{elsarticle}
\usepackage{amsmath,slashed,booktabs}
\usepackage{graphicx,graphics}
\usepackage{dcolumn}
\usepackage[hyperfootnotes=false]{hyperref}
\usepackage{xspace}
\usepackage{color}
\usepackage{balance}
\usepackage{multirow}
\usepackage{multicol}

\usepackage{fancyhdr}
\fancypagestyle{firstpage}{%
	
	\lhead{}
	\rhead{\small FERMILAB-PUB-22-443-V, LTH 1304, PSI-PR-22-13, ZU-TH 20/22}
}
\newcommand{\GeV}{\,\text{GeV}}

\newcommand{\fm}{\,\text{fm}}

\newcommand{\beq}{\begin{equation}}
\newcommand{\eeq}{\end{equation}}

\newcommand{\HVPexpref}{Bai:1999pk,Akhmetshin:2000ca,Akhmetshin:2000wv,Achasov:2000am,Bai:2001ct,Achasov:2002ud,Akhmetshin:2003zn,Aubert:2004kj,Aubert:2005eg,Aubert:2005cb,Aubert:2006jq,Aulchenko:2006na,Achasov:2006vp,Akhmetshin:2006wh,Akhmetshin:2006bx,Akhmetshin:2006sc,Aubert:2007ur,Aubert:2007ef,Aubert:2007uf,Aubert:2007ym,Akhmetshin:2008gz,Ambrosino:2008aa,Ablikim:2009ad,Aubert:2009ad,Ambrosino:2010bv,Lees:2011zi,Lees:2012cr,Lees:2012cj,Babusci:2012rp,Akhmetshin:2013xc,Lees:2013ebn,Lees:2013uta,Lees:2014xsh,Achasov:2014ncd,Aulchenko:2014vkn,Akhmetshin:2015ifg,Ablikim:2015orh,Shemyakin:2015cba,Anashin:2015woa,Achasov:2016bfr,Achasov:2016lbc,TheBaBar:2017aph,CMD-3:2017tgb,TheBaBar:2017vzo,Kozyrev:2017agm,Anastasi:2017eio,Achasov:2017vaq,Xiao:2017dqv,TheBaBar:2018vvb,Anashin:2018vdo,Achasov:2018ujw,Lees:2018dnv,CMD-3:2019ufp}

\allowdisplaybreaks[1]

\begin{document}

\renewcommand{\theequation}{\arabic{equation}}

\begin{frontmatter}
 
\title{Data-driven evaluations of Euclidean windows to scrutinize hadronic vacuum polarization}

\author[Bern]{G.~Colangelo}
\author[Illinois,Fermilab]{A.~X.~El-Khadra}
\author[Bern]{M.~Hoferichter}
\author[Manchester]{A.~Keshavarzi}
\author[Regensburg]{C.~Lehner}
\author[Zurich,PSI]{P.~Stoffer}
\author[Liverpool]{T.~Teubner}

\address[Bern]{Albert Einstein Center for Fundamental Physics, Institute for Theoretical Physics, University of Bern, Sidlerstrasse 5, 3012 Bern, Switzerland}
\address[Illinois]{Department of Physics and Illinois Center for Advanced Studies of the Universe, University of Illinois, Urbana, IL 61801, USA}
\address[Fermilab]{Particle Theory Department, Theory Division, Fermi National Accelerator Laboratory, Batavia, IL 60510, USA}
\address[Manchester]{Department of Physics and Astronomy, The University of Manchester, Manchester M13 9PL, United Kingdom}
\address[Regensburg]{Universit\"at Regensburg, Fakult\"at f\"ur Physik, Universit\"atsstra\ss e 31, 93040 Regensburg, Germany}
\address[Zurich]{Physik-Institut, Universit\"at Z\"urich,	Winterthurerstrasse 190, 8057 Z\"urich, Switzerland}
\address[PSI]{Paul Scherrer Institut, 5232 Villigen PSI, Switzerland}
\address[Liverpool]{Department of Mathematical Sciences, University of Liverpool, Liverpool, L69 3BX, United Kingdom}

\begin{abstract}
 In this paper, we discuss how windows in Euclidean time can be used to isolate the origin of potential conflicts between evaluations of the hadronic-vacuum-polarization (HVP) contribution to the anomalous magnetic moment of the muon in lattice QCD and from $e^+e^-\to\text{hadrons}$ cross-section data. We provide phenomenological comparison numbers evaluated from $e^+e^-\to\text{hadrons}$ data for the window quantities most commonly studied in lattice QCD, complete with the correlations among them. We discuss and evaluate modifications of window parameters that could be useful in dissecting the energy dependence of tensions in the HVP integral and emphasize that further optimizations require a precise knowledge of the full covariance matrix in lattice-QCD calculations as well.        
\end{abstract}

\end{frontmatter}

\thispagestyle{firstpage}

\section{Introduction}

Hadronic vacuum polarization (HVP) currently dominates the uncertainty in the Standard-Model prediction of the anomalous magnetic moment of the muon~\cite{Aoyama:2020ynm,Aoyama:2012wk,Aoyama:2019ryr,Czarnecki:2002nt,Gnendiger:2013pva,Davier:2017zfy,Keshavarzi:2018mgv,Colangelo:2018mtw,Hoferichter:2019gzf,Davier:2019can,Keshavarzi:2019abf,Kurz:2014wya,Melnikov:2003xd,Masjuan:2017tvw,Colangelo:2017fiz,Hoferichter:2018kwz,Gerardin:2019vio,Bijnens:2019ghy,Colangelo:2019uex,Blum:2019ugy,Colangelo:2014qya}
\beq
\label{amuSM}
a_\mu^\text{SM}=11\,659\,181.0(4.3)\times 10^{-10} \, , 
\eeq
quoted as the final recommendation in
Ref.~\cite{Aoyama:2020ynm} based on 
$e^+e^-\to\text{hadrons}$ cross-section data~\cite{Davier:2017zfy,Keshavarzi:2018mgv,Colangelo:2018mtw,Hoferichter:2019gzf,Davier:2019can,Keshavarzi:2019abf}
\beq
a_\mu^\text{HVP}\big|_{e^+e^-}=693.1(4.0)\times 10^{-10} \, .
\label{HVPee}
\eeq
Reference~\cite{Aoyama:2020ynm} also presented an average of lattice-QCD results available at the time~\cite{Chakraborty:2017tqp,Borsanyi:2017zdw,Blum:2018mom,Giusti:2019xct,Shintani:2019wai,FermilabLattice:2019ugu,Gerardin:2019rua,Aubin:2019usy,Giusti:2019hkz}
\beq
a_\mu^\text{HVP}\big|_\text{lattice}=711.6(18.4)\times 10^{-10} \, ,
\eeq
with a central value substantially higher than Eq.~\eqref{HVPee}, but consistent within uncertainties. It is the data-driven HVP evaluation entering Eq.~\eqref{amuSM} that,
compared to the current experimental world average~\cite{Muong-2:2006rrc,Muong-2:2021ojo,Muong-2:2021ovs,Muong-2:2021xzz,Muong-2:2021vma}
\beq
\label{exp}
a_\mu^\text{exp}=11\,659\,206.1(4.1)\times 10^{-10} \, ,
\eeq
produces a $4.2\sigma$ tension. 

Subsequently, BMWc published a lattice-QCD result for HVP~\cite{Borsanyi:2020mff} 
\beq
a_\mu^\text{HVP}\big|_\text{BMWc}=707.5(5.5)\times 10^{-10}
\eeq
that reduces the difference to $a_\mu^\text{exp}$ to $1.5\sigma$, while being in $2.1\sigma$ tension with the $e^+e^-$ determination of the HVP contribution. As first priority, it is essential that this result be confronted with independent lattice calculations at a similar level of precision~\cite{Colangelo:2022jxc}. This would be similar to Eq.~\eqref{HVPee} relying on a large number of data sets~\cite{\HVPexpref} and the quantification of tensions in the contributing database and analyses being reflected as an integral part of the quoted uncertainty estimate.

In response to Ref.~\cite{Borsanyi:2020mff} it has been emphasized that connections between $e^+e^-$ cross-section data, the HVP contribution, hadronic corrections to the running of the fine-structure constant $\alpha$, the global electroweak fit, and related low-energy parameters such as the pion charge radius can help trace the origin of the tension to specific energy ranges~\cite{Crivellin:2020zul,Keshavarzi:2020bfy,Malaescu:2020zuc,Colangelo:2020lcg}. Comparisons like this are particularly useful given that, e.g., the dispersion integral that determines the hadronic running involves a different energy weighting than $a_\mu^\text{HVP}$~\cite{Passera:2008jk}. More recently, a lattice-QCD evaluation of this running~\cite{Ce:2022eix} suggested a similar tension with $e^+e^-$ data as observed in Ref.~\cite{Borsanyi:2020mff}, further motivating more detailed scrutiny of the comparison between lattice QCD and the data-driven approach.
 
 \begin{figure*}[t]
	\centering
	\includegraphics[width=0.42\linewidth,clip]{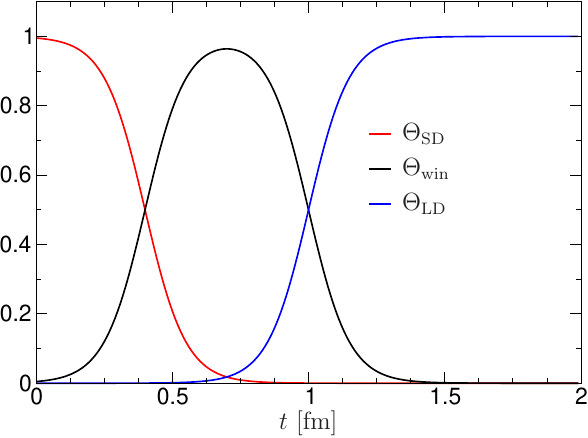}\qquad
	\includegraphics[width=0.42\linewidth,clip]{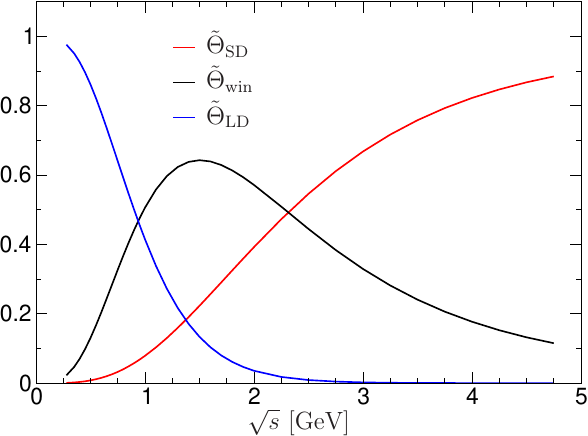}
	\caption{Short-distance, intermediate, and long-distance weight functions in Euclidean time (left), and their correspondence in center-of-mass energy (right).}
	\label{fig:kernels}
\end{figure*}

Another valuable tool for such comparisons is given by partial HVP integrals, the so-called window quantities, first suggested in Ref.~\cite{Blum:2018mom} and further developed in Ref.~\cite{Lehner:2020crt}.  These window quantities can be defined for fixed lattice spacing and have a well-defined continuum limit.  In the continuum theory a similar idea had been discussed before in Ref.~\cite{Bernecker:2011gh}. Adding a weight function to the HVP integral allows one to focus on certain regions in parameter space: either Euclidean time in the lattice approach or center-of-mass energy in phenomenology. In this paper, the possibilities regarding which window parameters are most useful for such comparisons are considered in detail and values for the same quantities are evaluated from $e^+e^-$ data and provided. Given that several lattice collaborations (FNAL/HPQCD/MILC and RBC/UKQCD) have started to implement blind analyses, it is well motivated to compile such phenomenological reference points to allow one to draw more immediate conclusions once new lattice results become available.  
 
In Sec.~\ref{sec:windows}, we provide such comparison numbers for the standard windows from Ref.~\cite{Blum:2018mom}, with $e^+e^-$ uncertainties treated in the same spirit as in Ref.~\cite{Aoyama:2020ynm}. In Sec.~\ref{sec:modified_windows}, we then consider a set of modified window quantities that should allow for a more detailed analysis of the energy dependence. The correlations among the different windows are also evaluated and included. Finally, we discuss the challenges in constructing optimized window observables to isolate the origin of potential conflicts between $e^+e^-$ data and lattice QCD.

 \section{Euclidean windows}
 \label{sec:windows}

 \begin{table}[t]
\renewcommand{\arraystretch}{1.3}
	\centering
	\scalebox{0.8}{
	\begin{tabular}{l c c c c}
	\toprule
	 & $a_\text{SD}^\text{HVP}$ & $a_\text{int}^\text{HVP}$ & $a_\text{LD}^\text{HVP}$ & $a_\text{total}^\text{HVP}$\\\midrule
	\multirow{2}{*}{All channels} & $68.4(5)$ & $229.4(1.4)$ & $395.1(2.4)$ & $693.0(3.9)$\\
	& $[9.9\%]$ & $[33.1\%]$ & $[57.0\%]$ & $[100\%]$\\
	\multirow{2}{*}{$2\pi$ below $1.0\GeV$} & $13.7(1)$ & $138.3(1.2)$ & $342.3(2.3)$ & $494.3(3.6)$\\
	& $[2.8\%]$ & $[28.0\%]$ & $[69.2\%]$ & $[100\%]$\\
	\multirow{2}{*}{$3\pi$ below $1.8\GeV$} & $2.5(1)$ & $18.5(4)$ & $25.3(6)$ & $46.4(1.0)$\\
	& $[5.5\%]$ & $[39.9\%]$ & $[54.6\%]$ & $[100\%]$\\\midrule
	White Paper~\cite{Aoyama:2020ynm} & -- & -- & -- & $693.1(4.0)$\\
	RBC/UKQCD~\cite{Blum:2018mom} & -- & $231.9(1.5)$ & -- & $715.4(18.7)$\\
	BMWc~\cite{Borsanyi:2020mff} & -- & $236.7(1.4)$ & -- & $707.5(5.5)$\\
	BMWc/KNT~\cite{Borsanyi:2020mff,Keshavarzi:2018mgv} & -- & $229.7(1.3)$ & -- & --\\
	Mainz/CLS~\cite{Ce:2022kxy} & -- & \!$237.30(1.46)$ & -- & --\\
	ETMC~\cite{Alexandrou:2022amy} & \!\!\!$69.33 (29)$\!\!\! & $235.0 (1.1) $ & -- & --\\
\bottomrule
	\end{tabular}
	}
	\caption{Window quantities for HVP, based on Refs.~\cite{Keshavarzi:2018mgv,Colangelo:2018mtw,Hoferichter:2019gzf,Keshavarzi:2019abf}, using the merging procedure from Ref.~\cite{Aoyama:2020ynm} and the window parameters~\eqref{window_standard} (for all channels, $2\pi$ below $1.0\GeV$, and $3\pi$ below $1.8\GeV$; in each case indicating the decomposition of the total in $\%$). Previous results from lattice QCD and phenomenology are shown for comparison where available (the quoted phenomenological evaluation of $a_\text{int}^\text{HVP}$ from Ref.~\cite{Borsanyi:2020mff} is based on Ref.~\cite{Keshavarzi:2018mgv}). We also include Refs.~\cite{Ce:2022kxy,Alexandrou:2022amy}, which appeared after the initial submission of our paper.
	All numbers in units of $10^{-10}$.}
	\label{tab:windows}
\end{table}

The master formula for the HVP contribution in the data-driven approach reads~\cite{Bouchiat:1961lbg,Brodsky:1967sr}
\begin{align}
\label{amu_HVP}
 a_\mu^\text{HVP}&=\bigg(\frac{\alpha m_\mu}{3\pi}\bigg)^2\int_{s_\text{thr}}^\infty ds \frac{\hat K(s)}{s^2}R_\text{had}(s) \, ,\notag\\
 R_\text{had}(s)&=\frac{3s}{4\pi\alpha^2}\sigma(e^+e^-\to\text{hadrons}(+\gamma)) \, ,
\end{align}
with kernel function 
\begin{align}
 \hat K(s) &= \frac{3s}{m_\mu^2} \Bigg[
		 \frac{x^2}{2} \big(2-x^2\big)+ \frac{1+x}{1-x} x^2 \log x \notag\\
		 &+ \frac{\big(1+x^2\big)(1+x)^2}{x^2}\bigg( \log(1+x) - x + \frac{x^2}{2} \bigg)  \Bigg] \, ,\notag\\
		x &= \frac{1-\sigma_\mu(s)}{1+\sigma_\mu(s)} \, ,\qquad \sigma_\mu(s)=\sqrt{1-\frac{4m_\mu^2}{s}} \, .
\end{align}
The integration threshold takes the value $s_\text{thr}=M_{\pi^0}^2$, since the $\pi^0\gamma$ channel is included, by convention, in the photon-inclusive cross section (in the same way, final-state radiation is included, in particular in the $2\pi$ and $3\pi$ channels below). In lattice QCD, most collaborations employ the time-momentum representation~\cite{Lautrup:1971jf,Blum:2002ii,Bernecker:2011gh}
\begin{figure*}[t]
	\centering
	\includegraphics[width=0.42\linewidth,clip]{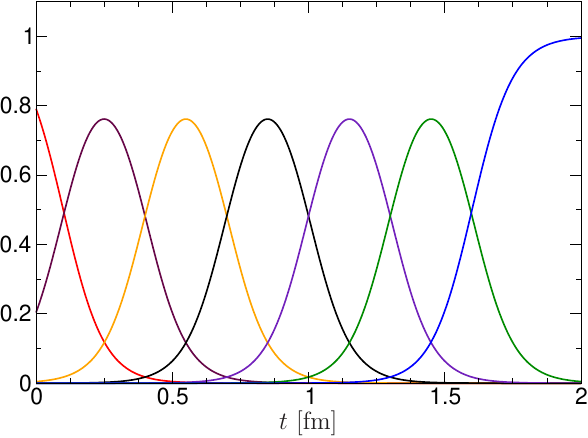}\qquad
	\includegraphics[width=0.42\linewidth,clip]{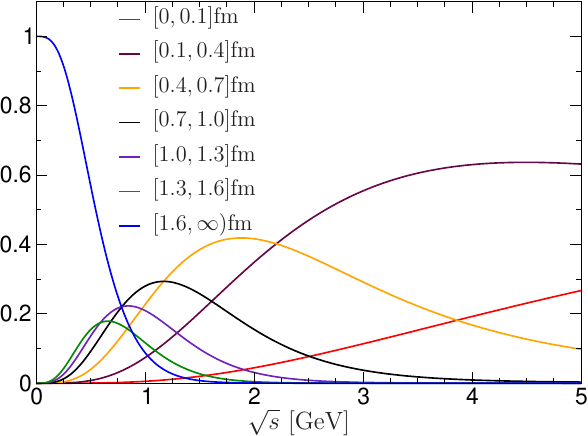}
	\caption{Analog of Fig.~\ref{fig:kernels} for a finer decomposition ($\Delta=0.15\fm$). The colors match the corresponding weight functions in Euclidean time (left) and center-of-mass energy (right).}
	\label{fig:kernels_fine}
\end{figure*} 
\begin{align}
 a_\mu^\text{HVP}&=\bigg(\frac{\alpha}{\pi}\bigg)^2\int_{0}^\infty dt\, \tilde K(t) G(t) \, ,
 \label{timemomentum}
\end{align}
with another known kernel function $\tilde K(t)$ and $G(t)$ given by the correlator of two electromagnetic currents $j_\mu^\text{em}$
\begin{align}
 G(t)&=-\frac{a^3}{3}\sum_{k=1}^3\sum_{\mathbf x}G_{kk}(t,{\mathbf x}) \, ,\notag\\
 G_{\mu\nu}(x)&=\langle 0|j_\mu^\text{em}(x)j_\nu^\text{em}(0)|0\rangle \, ,
\end{align}
with the lattice spacing taken to the limit $a\to 0$. Windows in Euclidean time are defined by an additional weight function in Eq.~\eqref{timemomentum}.
The ones proposed in Ref.~\cite{Blum:2018mom}
\begin{align}
\Theta_\text{SD}(t)&=1-\Theta(t,t_0,\Delta) \, ,\notag\\
 \Theta_\text{win}(t)&=\Theta(t,t_0,\Delta)-\Theta(t,t_1,\Delta) \, ,\notag\\
 \Theta_\text{LD}(t)&=\Theta(t,t_1,\Delta) \, ,\notag\\
 \Theta(t,t',\Delta)&=\frac{1}{2}\bigg(1+\tanh\frac{t-t'}{\Delta}\bigg) \,,
\end{align}
were designed to separate short-distance, intermediate, and long-distance contributions, respectively, with parameters 
\beq
t_0=0.4\fm\,,\qquad t_1=1.0\fm\,,\qquad \Delta=0.15\fm\,.
\label{window_standard}
\eeq
The isospin-symmetric quark-connected light-quark contribution of the intermediate window for these parameters has now been calculated by several lattice collaborations at high precision~\cite{Blum:2018mom,Aubin:2019usy,Borsanyi:2020mff,Lehner:2020crt,Giusti:2021dvd,Wang:2022lkq,Aubin:2022hgm}. Some collaborations have also computed the quark-disconnected and isospin-breaking corrections~\cite{Blum:2018mom,Borsanyi:2020mff}.
The corresponding weight functions $\tilde \Theta(s)$ in Eq.~\eqref{amu_HVP} are obtained as
\begin{align}
 \tilde \Theta(s)&=\frac{3 s^{5/2}}{8m_\mu^4 \hat K(s)}\int_0^\infty dt\, \Theta(t)e^{-t\sqrt{s}} \int_0^\infty ds'\, w\bigg(\frac{s'}{m_\mu^2}\bigg)\notag\\
 &\times\Bigg(t^2-\frac{4}{s'}\sin^2\frac{t\sqrt{s'} }{2}\Bigg) \, ,\notag\\
 w(r)&=\frac{\Big[r+2-\sqrt{r(r+4)}\Big]^2}{\sqrt{r(r+4)}} \, .
\end{align}

Results for the window parameters~\eqref{window_standard} are collected in Table~\ref{tab:windows}, including comparison numbers from $e^+e^-$ data obtained from Refs.~\cite{Keshavarzi:2018mgv,Colangelo:2018mtw,Hoferichter:2019gzf,Keshavarzi:2019abf} using the merging procedure from Ref.~\cite{Aoyama:2020ynm}. We have not included new data~\cite{SND:2020nwa,BESIII:2019gjz,BABAR:2021cde}  that became available after these references nor Refs.~\cite{Hoid:2020xjs,Stamen:2022uqh} for the $\pi^0\gamma$ and $\bar K K$ channel, respectively, given that the overall impact will be small and subtleties in the inclusion into global analyses first need to be assessed in each method separately. The $e^+e^-\to 2\pi$ data from Ref.~\cite{Ablikim:2015orh} (including the corrected covariance matrix), however, have been added to the analysis of Ref.~\cite{Colangelo:2018mtw}, which ensures a realistic estimate of the systematic tension between the BaBar and KLOE data close to the one included in Refs.~\cite{Aoyama:2020ynm,Davier:2019can}. With these numbers, the global $2.1\sigma$ tension between Ref.~\cite{Borsanyi:2020mff} and $e^+e^-$ data increases to $3.7\sigma$ in the intermediate window. The result from Ref.~\cite{Blum:2018mom} lies $1.2\sigma$ above $e^+e^-$ data and $2.3\sigma$ below Ref.~\cite{Borsanyi:2020mff}.  

Table~\ref{tab:windows} also shows the decomposition for the two leading hadronic channels, indicating how their contributions are distributed over the three windows, as well as the 
extent to which the $2\pi$ channel dominates the long-distance window. As can be seen from Fig.~\ref{fig:kernels}, the sharp separation into short-distance, intermediate, and long-distance weight functions in Euclidean time becomes far more ambiguous in center-of-mass energy, with significant overlap of the windows and a long tail of the intermediate weight function. Accordingly, this window still receives the dominant contribution ($\sim60\%$) from the $2\pi$ channel (to be compared to $71\%$ for the total HVP), but a significant part comes from higher-multiplicity channels and the inclusive region above $(1.8\text{--}2)\GeV$. In contrast, the long-distance window is strongly dominated by the $2\pi$ ($87\%$) and $3\pi$ ($6\%$) channels.

\begin{table}[t]
\renewcommand{\arraystretch}{1.3}
	\centering
	\scalebox{0.85}{
	\begin{tabular}{l c c c}
	\toprule
	& $2\pi$ $\leq 1.0\GeV$ &
	$3\pi$ $\leq 1.8\GeV$ & All channels\\\midrule
	$[0,0.1]$fm & $0.83(0)(1)$ & $0.18(0)(0)$ & $11.43(9)$\\
	$[0.1,0.4]$fm & $12.89(5)(11)$ & $2.37(4)(2)$ & $57.01(41)$\\
	$[0.4,0.7]$fm & $51.02(19)(41)$ & $7.69(14)(6)$ & $102.54(62)$\\
	$[0.7,1.0]$fm & $87.28(31)(65)$ & $10.82(21)(7)$ & $126.89(79)$\\
	$[1.0,1.3]$fm & $95.31(34)(65)$ & $9.84(20)(5)$ & $120.51(77)$\\
	$[1.3,1.6]$fm & $80.88(30)(50)$ & $6.97(15)(2)$ & $95.01(60)$\\
	$[1.6,\infty)$fm & $166.08(80)(69)$ & $8.53(19)(2)$ & $179.64(1.08)$\\\midrule
	Total & $494.30(1.90)(3.00)$ & $46.39(94)(24)$ & $693.02(3.86)$\\
\bottomrule
	\end{tabular}
	}
	\caption{Window quantities for HVP, based on Refs.~\cite{Keshavarzi:2018mgv,Colangelo:2018mtw,Hoferichter:2019gzf,Keshavarzi:2019abf}, using the merging procedure from Ref.~\cite{Aoyama:2020ynm} and the window parameters shown in Fig.~\ref{fig:kernels_fine}. The first and second errors for the $2\pi$ and $3\pi$ channels refer to the experimental and additional systematic uncertainties, respectively, as described in the main text. 
	All numbers in units of $10^{-10}$.}
	\label{tab:windows_ext}
\end{table} 

\begin{figure*}[t]
	\centering
	\includegraphics[width=0.42\linewidth,clip]{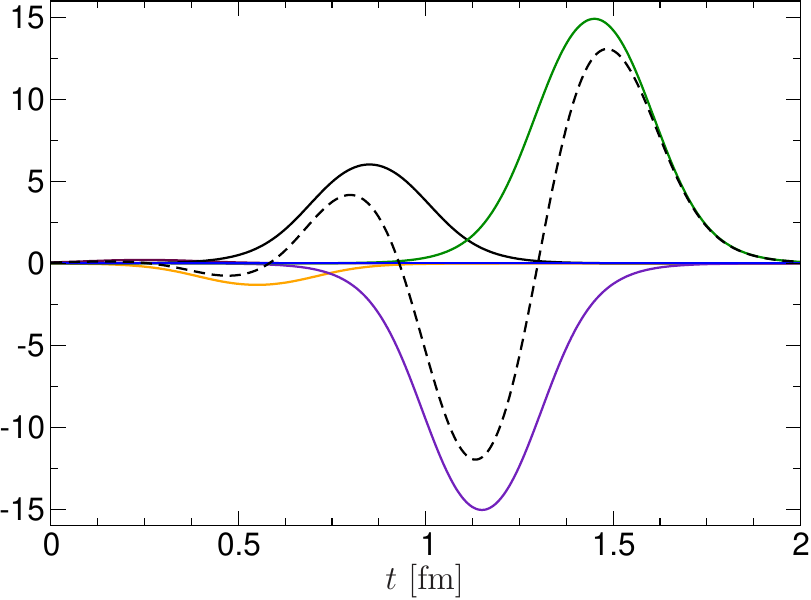}\qquad
	\includegraphics[width=0.42\linewidth,clip]{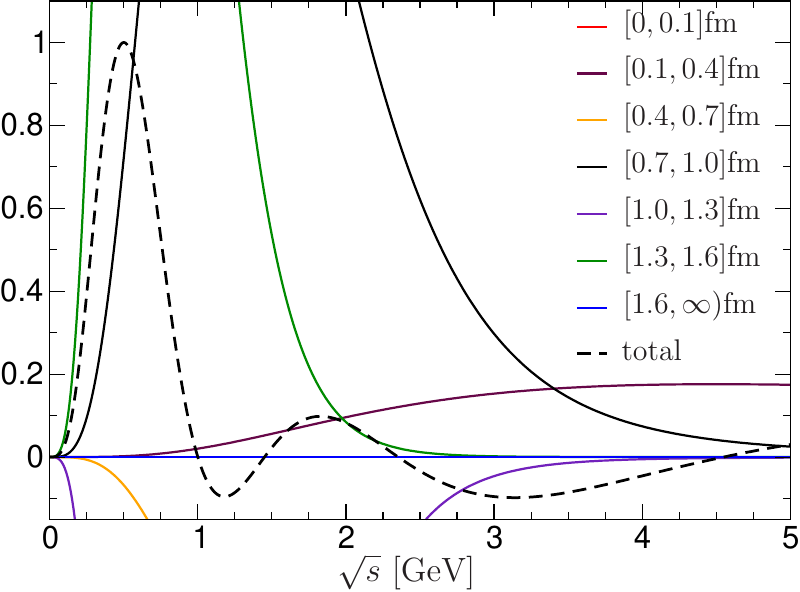}
	\caption{Example for a linear combination of our modified windows. The better localization in center-of-mass energy (right) requires severe cancellations between the different windows, leading to strong oscillations in Euclidean time (left). Note the different scales of the two plots. The color-coding is as in Fig.~\ref{fig:kernels_fine}.}
	\label{fig:kernels_cancellations}
\end{figure*} 
 
 \section{Modified window parameters}
 \label{sec:modified_windows}
 
 More detailed information on the energy dependence can be obtained starting from a finer decomposition in Euclidean time~\cite{Lehner:2020crt}. In Fig.~\ref{fig:kernels_fine} we show a decomposition in which the intermediate window is cut in half and windows of the same time difference $0.3\fm$ (with $\Delta=0.15\fm$) are added in both directions. The overlap of the weight functions in center-of-mass energy is substantial, but the main support of the windows still differs sufficiently such that conclusions on the energy dependence of potential differences between lattice and phenomenological evaluations should be possible. In particular, any trends identified in the three-window scenario from Fig.~\ref{fig:kernels} could be scrutinized in this more detailed decomposition. For example, if a more significant tension in the intermediate window compared to the global integral were corroborated, the consideration of the refined windows proposed here should allow one to better locate the origin of the differences. 
 
 \begin{table}[t]
\renewcommand{\arraystretch}{1.3}
	\centering
	\scalebox{0.79}{
	\begin{tabular}{c c c c c c c}
	\toprule
	$1$ &	$0.901521$ & 	$0.471482$ &	$0.194165$ &	$0.120959$ &	$0.099851$ &	$0.076151$\\
&	$1$ &	$0.758581$	& $0.469941$ &	$0.364565$ &	$0.324456$ &	$0.256454$\\
& & $1$ & $0.909295$ &	$0.827037$ &	$0.780636$ &	$0.641164$\\
& & & $1$	& $0.982477$ &	$0.958442$ &	$0.813646$\\
& & & & $1$	& $0.993619$ &	$0.871170$\\
& & & & & $1$ &	$0.911101$\\
& & & & & & $1$\\
\bottomrule
	\end{tabular}
	}
	\caption{Final correlations among the window quantities given in Table~\ref{tab:windows_ext} (all channels). The correlations among the standard windows can be reconstructed from this covariance matrix together with Table~\ref{tab:windows_ext}: $\rho_\text{SD, win}=0.566$, $\rho_\text{SD, LD}=0.280$, $\rho_\text{win, LD}=0.872$.}
	\label{tab:windows_ext_corr}
\end{table}
 
 Our results for these modified windows are given in Table~\ref{tab:windows_ext}. In analogy to the breakdown in Table~\ref{tab:windows}, we provide the separate results for the $2\pi$ and $3\pi$ channels from center-of-mass energies below $1.0\GeV$ and $1.8\GeV$, respectively. In both cases, the uncertainties already include systematic effects as prescribed by the merging procedure from Ref.~\cite{Aoyama:2020ynm}. The central values for the $2\pi$ and $3\pi$ channels are determined as the average of Ref.~\cite{Keshavarzi:2019abf} with Refs.~\cite{Colangelo:2018mtw,Hoferichter:2019gzf}. All experimental errors are carried over from Ref.~\cite{Keshavarzi:2019abf}, as are the remaining contributions to the HVP integral. The systematic uncertainty in the $3\pi$ channel is given by half the difference between Refs.~\cite{Keshavarzi:2019abf,Hoferichter:2019gzf}, whereas in the $2\pi$ channel is obtained as half the difference between fits without the BaBar and without the KLOE data as in the implementation of Ref.~\cite{Colangelo:2018mtw} (as the effect is larger than the corresponding one in Ref.~\cite{Keshavarzi:2019abf}). In comparison to the full merging procedure from Ref.~\cite{Aoyama:2020ynm}, the main omissions in the present estimate concern the effect of interchannel correlations~\cite{Davier:2017zfy,Davier:2019can} and an additional systematic uncertainty assigned for the inclusive region. Numerically, however, these effects are subleading and compensated by a slightly larger value for the BaBar/KLOE tension, so that our final result for the total HVP contribution comes out very close to Ref.~\cite{Aoyama:2020ynm}. 
 
In view of the overlap of the weight functions, the uncertainties of the window quantities derived in this way display significant correlations. To facilitate usage of our results (e.g., for the construction of linear combinations), the final correlations are provided in Table~\ref{tab:windows_ext_corr}. The covariance matrix is derived starting from Ref.~\cite{Keshavarzi:2019abf} for the experimental uncertainties, to which the covariance matrices corresponding to the systematic uncertainties for the $2\pi$ and $3\pi$ channels (each $100\%$ correlated among the windows) are added. Once these correlations are taken into account, sums of the modified windows reproduce the standard ones as given in Table~\ref{tab:windows}, as well as the total HVP contribution. More details on the covariance matrices are provided in~\ref{app:covariance}.

From the perspective of lattice-QCD calculations, the additional windows discussed in this work will exhibit a different balance of statistical and systematic uncertainties compared to the standard $[0.4,1.0]\fm$ window. At shorter distances, discretization errors may be enhanced, while at longer distances both statistical uncertainties and the size of needed finite-volume corrections grow. As pointed out in Ref.~\cite{Aubin:2022hgm}, chiral perturbation theory ($\chi$PT) describes the long-distance windows much better: lattice calculations that rely on the applicability of $\chi$PT~\cite{Sharpe:2004is,Bijnens:2017esv,Aubin:2020scy,Colangelo:2021moe} need to take this into account as well.
 
The splitting of the total integral over Euclidean time into more window quantities has a predominantly illustrative purpose. Depending on the specific goal one has in mind, other window quantities than those discussed here may be better suited to single out specific energy regions or contributions. This must be studied on a case-by-case basis and must take into account aspects of the lattice-QCD calculations to which one wants to compare. This can be done also by building appropriate linear combinations of the windows discussed here. 

To illustrate this, we show an example in Fig.~\ref{fig:kernels_cancellations} that is built to have a significant localization in $\sqrt{s}$ (another example, together with some details on how to build such linear combinations, is provided in~\ref{app:combinations}).  The example shows that, to achieve this, strong cancellations between different windows must occur, which tends to enhance uncertainties. This is a manifestation of the corresponding ill-posed inverse Laplace transform. In this context, as recently also discussed in Ref.~\cite{DeGrand:2022lmc}, precise knowledge of the covariance matrix of the windows within a lattice-QCD calculation is needed. The different setups regarding statistical sampling techniques, as well as the treatment of lattice systematics, make it difficult to provide a generally optimal recommendation. We encourage lattice collaborations to consider computing the additional modified windows and make the covariance matrices available to facilitate such optimizations.

 \section{Conclusions}
 \label{sec:conclusions}
 
 In this work, we discussed the role of Euclidean window quantities to scrutinize the origin of potential conflicts between evaluations of the HVP contribution to the anomalous magnetic moment of the muon in lattice QCD and from $e^+e^-\to\text{hadrons}$ data. As a first step, we provided phenomenological comparison numbers for the standard windows typically considered in lattice-QCD calculations, following Ref.~\cite{Aoyama:2020ynm} to account for tensions in the database. We then extended this setup in a minimal way that allows for more detailed studies of different energy regions, by dividing the intermediate window in half and adding commensurate Euclidean windows in either direction. Again, we provided the corresponding comparison numbers for these windows and the correlations among them, which allows building and studying general linear combinations.  
 
While this choice of windows, as illustrated in Fig.~\ref{fig:kernels_fine}, can be considered a minimal extension, it is by no means unique. Depending on forthcoming lattice-QCD results, other variants might prove more useful in isolating phenomenological tensions or minimizing specific sources of uncertainty. Such modifications can be implemented following the procedure established here. However, optimizing window observables also requires precise knowledge of the covariance matrices in lattice-QCD calculations. We expect that the comparison to data-driven results for Euclidean window quantities should prove valuable for understanding the origin of differences between phenomenological and lattice-QCD HVP evaluations in the future.

\section*{Acknowledgments}

 Special thanks are extended to D.~Nomura for his collaboration with A.~Keshavarzi and T.~Teubner in producing the compilation of hadronic cross-section data from Ref.~\cite{Keshavarzi:2019abf} used in this work.
Financial support from the SNSF (Project Nos.\ 200020\_175791, PCEFP2\_181117, and PCEFP2\_194272), the STFC Consolidated Grants ST/S000925/1 and ST/T000988/1, the European Union’s Horizon 2020 research and innovation programme under the Marie Sk\l{}odowska-Curie grant agreement No.\ 858199 (INTENSE), and the US Department of Energy, Office of Science, Office of High energy Physics under Award No.\ DE-SC0015655 is gratefully acknowledged.
 
 \appendix

 \section{Covariance matrices}
 \label{app:covariance}
 
 In this appendix, we provide additional details on the correlations of the window quantities studied in Sec.~\ref{sec:modified_windows}. The uncertainties in each window are already separated into experimental and additional systematic uncertainties for the $2\pi$ and $3\pi$ channels in Table~\ref{tab:windows_ext}. These systematic errors are added in quadrature to the total experimental error from Ref.~\cite{Keshavarzi:2019abf} to obtain the final uncertainty. In a similar fashion, the full covariance matrix is obtained by summing the experimental covariance matrix with the additional systematic covariance matrices. Since the entries of the latter are $100\%$ correlated among all windows, they follow in a straightforward way from Table~\ref{tab:windows_ext}. The experimental correlations are given explicitly in Table~\ref{tab:windows_ext_corr_exp}. 
 
 \begin{table}[b]
\renewcommand{\arraystretch}{1.3}
	\centering
	\scalebox{0.79}{
	\begin{tabular}{c c c c c c c}
	\toprule
 $1$ &	$0.916686$ &	$0.563435 $&$	0.228185 $&$	0.102148 $&$	0.062811 $&$	0.034410 $\\
	 &$1$ &	$0.807352 $&$	0.456399	$&$ 0.269911 $&$	0.193444 $&$	0.117589$\\
	& & $1$ &	$0.853005 $&$	0.665217 $&$	0.554058	$&$ 0.379963$\\
	& & &$1$	&$0.946031$&$	0.873679$&$	0.660724$\\
	& & & & $1$	&$0.981179	$&$0.803188$\\
	& & & & & $1$	&$0.882686$\\
& & & & & &$1$\\
\bottomrule
\end{tabular}
}
	\caption{Correlations of the experimental uncertainties among the window quantities given in Table~\ref{tab:windows_ext} (all channels), derived from Ref.~\cite{Keshavarzi:2019abf}.}
	\label{tab:windows_ext_corr_exp}
\end{table}

\begin{table*}[t!]
\renewcommand{\arraystretch}{1.3}
	\centering
	\scalebox{0.85}{
	\begin{tabular}{c c c c c c c}
	\toprule 
1	& 0.9933525066 &	0.9688971356&	0.9392576164 &	0.9096626614	& 0.8738177074	& 0.6793973761\\
0.9998213506502&	1	&0.9906916562&	0.9708649531&	0.9463823504&	0.9128301789&	0.7145351809\\
0.9986149451513	&0.9994301486018&	1	&0.9938807916&	0.9782815834&	0.9507022306&	0.7574812008\\
0.9955029998681&	0.9971080510323&	0.9991018537292&	1&	0.9946021546&	0.9760842026&	0.8011055545\\
0.9899396198572&	0.9924077721485&	0.9959702858346&	0.9988699236489&	1	&0.9930330675&	0.8492090182\\
0.9815319618100&	0.9848948727493&	0.9900980686032	&0.9951170152493&	0.9986752122804&	1	&0.9008970236
\\
0.9463341811594&	0.9518235165415&	0.9610079563233&	0.9713226940780&	0.9811611726780&	0.9895962983862&	1\\
\bottomrule
\end{tabular}
}
	\caption{Correlations in the $2\pi$ channel, for Ref.~\cite{Keshavarzi:2019abf} (upper triangle) and Ref.~\cite{Colangelo:2018mtw} (lower triangle). The latter do not include the systematic uncertainties of the dispersive representation but reflect the correlations of the fit parameters.}
	\label{tab:windows_ext_corr_2pi}
\end{table*}

\begin{table*}[t!]
\renewcommand{\arraystretch}{1.3}
	\centering
	\scalebox{0.85}{
	\begin{tabular}{c c c c c c c}
	\toprule 
1 &	0.992021647	&0.959302888&	0.920188802&	0.891183027&	0.871778808&	0.850332510 \\
0.99329507639	& 1 &	0.987158146&	0.961679561&	0.939999409&	0.924458447&	0.906262339\\
0.96246362304&	0.98730879253	& 1 &	0.992976286&	0.981747430&	0.972035048&	0.959105240\\
0.92038302321&	0.95889552857&	0.99160878168	& 1 &	0.997253863	&0.992586218&	0.984572290\\
0.88602301922&	0.93228208694&	0.97700793206&	0.99622025094	& 1 &	0.998822258&	0.994476746\\
0.86183265437&	0.91217905294&	0.96368319178&	0.98949209484&	0.99824954875	& 1 &	0.998317290\\
0.83561042158&	0.88912843415&	0.94642367704&	0.97843094492&	0.99226649272&	0.99781891435	& 1\\
\bottomrule
\end{tabular}
}
	\caption{Correlations in the $3\pi$ channel, for Ref.~\cite{Keshavarzi:2019abf} (upper triangle) and Ref.~\cite{Hoferichter:2019gzf} (lower triangle). The latter do not include the systematic uncertainties of the dispersive representation but reflect the correlations of the fit parameters.}
	\label{tab:windows_ext_corr_3pi}
\end{table*}

Finally, we also provide the correlations for the $2\pi$ and $3\pi$ channels in Table~\ref{tab:windows_ext_corr_2pi} and Table~\ref{tab:windows_ext_corr_3pi}, respectively. For the latter, the results from the direct integration of the data~\cite{Keshavarzi:2019abf} and the dispersive representation~\cite{Hoferichter:2019gzf} are very close, ultimately because the $3\pi$ cross section is strongly dominated by the $\omega$ and $\phi$ resonances. In the $2\pi$ channel, bigger differences are observed between the two methodologies as, in this case, the functional form imposed in the dispersive representation gives rise to strong correlations across the entire low-energy region. 

\section{Linear combinations}
\label{app:combinations}
 
In this appendix, we discuss examples of linear combinations of the modified window quantities presented in this paper. Denoting the weight functions of the seven windows by $\Theta_i(t)$, $i=1,\ldots,7$ (and the corresponding weight functions in squared center-of-mass energy by $\tilde\Theta_i(s)$), Fig.~\ref{fig:kernels_cancellations} shows the linear combination
\beq
	\Theta_\text{comb}(t) = \sum_{i=1}^7 a_i \Theta_i(t)\, , \qquad \tilde\Theta_\text{comb}(s) = \sum_{i=1}^7 a_i \tilde\Theta_i(s)\,,
\eeq
with coefficients $a_i$ given in the first row of Table~\ref{tab:windows_comb}. The momentum-space weight function has a maximum at $\sqrt{s} = 0.5 \GeV$ and stays in magnitude below $10\%$ of the peak value for $\sqrt{s} > 1 \GeV$. In Table~\ref{tab:windows_comb}, we list the contributions of the different channels to this linear combination, where the uncertainties have been propagated with the provided correlations. The localization at small energies leads to a dominance of the $2\pi$ channel that surpasses the one of the $[1.6,\infty)\fm$ Euclidean window. However, as shown in Fig.~\ref{fig:kernels_cancellations}, this comes at the expense of strong cancellations between the different windows, leading to a large oscillation amplitude in the Euclidean-time weight function, which will enhance the uncertainties of lattice-QCD calculations.

\begin{figure*}[t]
	\centering
	\includegraphics[width=0.42\linewidth,clip]{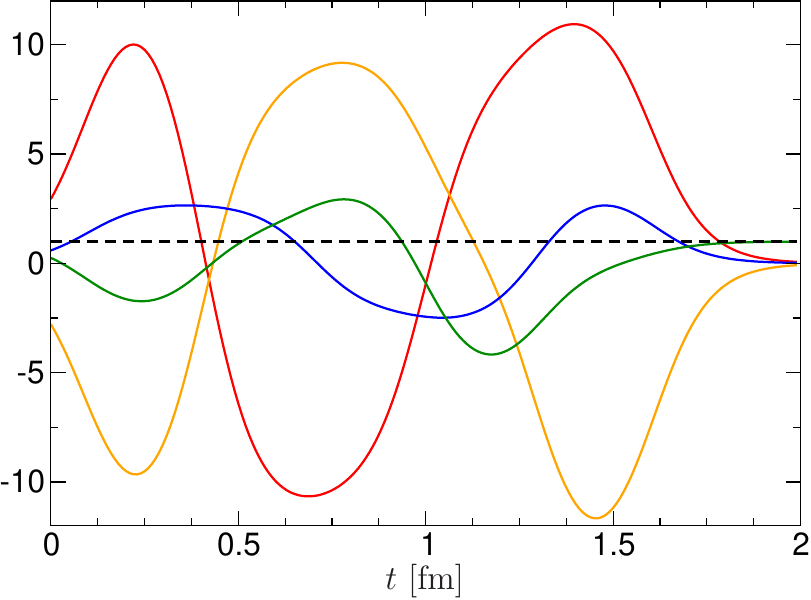}\qquad
	\includegraphics[width=0.42\linewidth,clip]{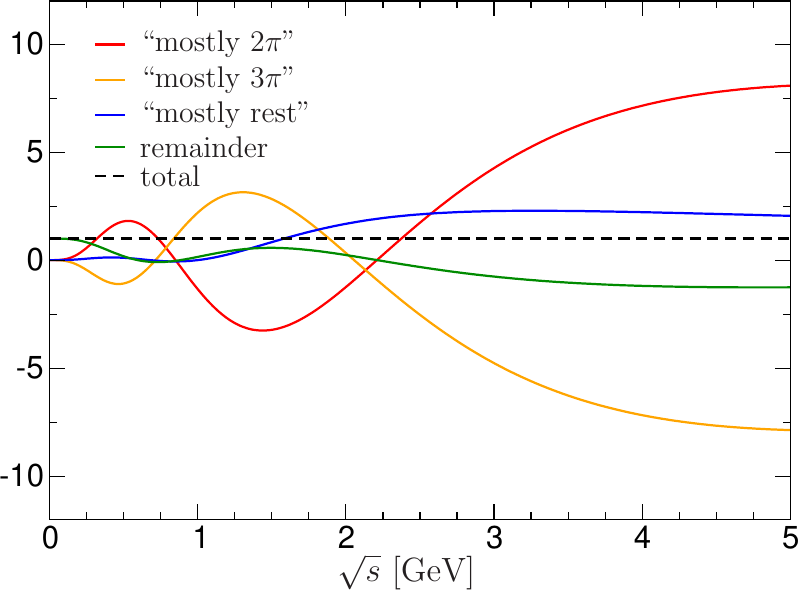}
	\caption{Examples for linear combinations of our modified windows that are dominated by contributions from different intermediate states. Although the combinations are chosen to minimize the amplitude of the Euclidean-time weight functions (left), one still observes rather strong oscillations, whereas the enhancement of cross-section uncertainties due to the center-of-mass energy weight functions (right) remains moderate.}
	\label{fig:channel_extraction}
\end{figure*}

As an alternative strategy to weight functions that are localized in center-of-mass energy, one could try to alleviate the problems connected with the inverse Laplace transform by constructing different linear combinations that are dominated by individual hadronic channels instead of localized energy regions. The second panel of Table~\ref{tab:windows_comb} and Fig.~\ref{fig:channel_extraction} show linear combinations of the seven windows that are defined as follows: the combinations ``mostly $2\pi$,'' ``mostly $3\pi$,'' and ``mostly rest'' reproduce the HVP contributions of the $2\pi$, $3\pi$, and remaining channels, respectively, from linear combinations of windows $2$ to $6$ only (windows $1$ and $7$ are expected to be hampered most by lattice-QCD uncertainties). Two free parameters in each combination are tuned to minimize the amplitude of the Euclidean-time weight functions. The last combination sums with the first three combinations to unit weight.

While in principle these combinations allow one to focus on contributions from individual channels, the oscillations in the Euclidean-time weight functions are still large, although somewhat reduced compared to the strategy of localized weight functions (rescaling the weight $\Theta_\text{comb}$ so that its $2\pi$ component matches the full $2\pi$ contribution leads to an amplitude twice as large as the one of the ``mostly $2\pi$'' Euclidean-time weight function). Note that these linear combinations mainly serve as illustrations but choosing different energy cutoffs for the $2\pi$ and $3\pi$ channels should not change the picture qualitatively. Future updates of the cross-section data would lead to admixtures of the different channels (see Table~\ref{tab:windows_comb} for the current uncertainties) unless the combinations are adjusted. Finally, the free parameters of the combinations could be tuned to minimize lattice-QCD uncertainties. Whether such a strategy could disentangle contributions from different channels in realistic lattice-QCD simulations depends on the covariance matrices, which thus provides further motivation to make both the window quantities and their correlations available.

\begin{table*}[t]
\renewcommand{\arraystretch}{1.3}
	\centering
	\scalebox{0.85}{
	\begin{tabular}{l c c c c c c c c c c c}
	\toprule
	& $a_1$ & $a_2$ & $a_3$ & $a_4$ & $a_5$ & $a_6$ & $a_7$ & $\;$
	& $2\pi$ $\leq 1.0\GeV$ &
	$3\pi$ $\leq 1.8\GeV$ & All channels\\
	\midrule
	\multirow{2}{*}{$\Theta_\text{comb}$}
					&	\multirow{2}{*}{$0$}	&	\multirow{2}{*}{$0.276$}	&	\multirow{2}{*}{$-1.719$}	&	\multirow{2}{*}{$7.918$}	&	\multirow{2}{*}{$-19.743$}	&	\multirow{2}{*}{$19.579$}	&	\multirow{2}{*}{$0$}	&& $308.78(1.33)(1.36)$ & $15.30(37)(0)$ & $325.15(1.93)$\\
					&&&&&&&&		& $[95.0\%]$ & $[4.7\%]$ & $[100\%]$\\
	\midrule
	``mostly $2\pi$''		&	$0$	&	$14.698$		&	$-11.994$		&	$-10.961$		&	$8.945$	&	$12.622$		&	$0$	&& $494.29(2.19)(1.67)$ & $0.00(48)(43)$ & $494.29(4.41)$\\
	``mostly $3\pi$''		&	$0$	&	$-13.847$		&	$8.657$		&	$10.177$		&	$1.081$	&	$-15.510$		&	$0$	&& $0.00(1.01)(1.63)$ & $46.40(78)(58)$ & $46.40(4.11)$\\
	``mostly rest''		&	$0$	&	$2.838$		&	$2.709$		&	$-2.308$		&	$-3.002$	&	$3.866$		&	$0$	&& $0.00(11)(12)$ & $0.00(6)(3)$ & $152.31(1.61)$\\
	remainder			& 	$1$	&	$-2.689$		&	$1.629$		&	$4.091$		&	$-6.025$	&	$0.022$		&	$1$	&& $0.00(32)(18)$ & $0.00(7)(6)$ & $0.00(68)$\\
	\bottomrule
	\end{tabular}
	}
	\caption{Linear combinations of window quantities for HVP, based on Refs.~\cite{Keshavarzi:2018mgv,Colangelo:2018mtw,Hoferichter:2019gzf,Keshavarzi:2019abf}, using the merging procedure from Ref.~\cite{Aoyama:2020ynm}. The row labeled by $\Theta_\text{comb}$ corresponds to Fig.~\ref{fig:kernels_cancellations}, the second panel gives the combinations shown in Fig.~\ref{fig:channel_extraction}. The first and second errors for the $2\pi$ and $3\pi$ channels refer to the experimental and additional systematic uncertainties, respectively, as described in the main text. 
	The numbers of the last three columns are in units of $10^{-10}$. Due to the large correlations, the rounded input from Table~\ref{tab:windows_ext} will lead to small deviations from the given numbers.}
	\label{tab:windows_comb}
\end{table*} 

\bibliographystyle{apsrev4-1_mod}
\balance
\biboptions{sort&compress}
\bibliography{AMM}

\end{document}